# MODES OF KINK MOTION ON DISLOCATIONS IN SEMICONDUCTORS

Yu. L. Iunin and V. I. Nikitenko

Institute of Solid State Physics, Russian Academy of Sciences, Chernogolovka,
Moscow district, 142432, e-mail: Iunin@issp.ac.ru

*Keywords*: Dislocations mobility; Elemental semiconductors; Semiconductor alloys

*Abstract.* Analysis is given of the changes of dislocation motion modes with stress and temperature variation. Different regimes of dislocation kink pair formation and spreading (motion in the random potential, in the field of random forces, the quasi-localization) are considered. Discrepancies are discussed between the theory and experimental data on dislocation velocities.

Introduced in 1953 concept of kinks [1] has become one of the most attractive in the fundamental physics. It allowed one to describe the microscopic mechanisms of generation of the dislocations, domain walls, Bloch lines etc. The theory of dislocation motion in the Peierls potential relief [2–5] has been developed basing on the kink idea. In the present work we compare predictions of the theory of dislocation motion in the deep potential relief with the experimental data on dislocation and kink dynamics obtained with Si, Ge, and SiGe single crystals.

## Modes of dislocation motion in the Peierls relief. Theory

According to a theory, several critical shear stress values exist, under which the dislocation velocity dependence on stress is qualitatively changed. One of them is the Peierls stress $\tau_P = (1/b)\left[\partial W(y)/\partial y\right]_{max}$ separating modes of the viscous ($\tau > \tau_P$) and the thermoactivated ($\tau \leq \tau_P$) dislocation motion. Here $W(y)$ is the dependence of dislocation energy on its position in the glide plane (Fig. 1a), $b$ is the magnitude of the Burgers vector of a dislocation.

In the first mentioned above mode dislocation mobility is determined by various mechanisms of the energy transmission from the dislocation to elementary excitations of the crystal lattice. In this regime dislocation velocity $V$ is proportional to the acting force $V=\tau b/B(T)$, where $B(T)$ is the dynamical drag increasing with the temperature $T$. So $V$ decreases with temperature in this regime. With $\tau \leq \tau_P$ dislocation motion occurs by the thermoactivated nucleation and spreading of the kink pairs. Unlike the viscous motion, in the thermoactivated mode dislocation velocity increases with a temperature $V\sim\exp(-U/kT)$. Here $U$ denotes the activation energy; $k$ is the Boltzman constant.

Thermoactivated dislocation motion is determined by the probability (per unit time per unit length of a dislocation) of the nucleation of a kink pair of critical size stable to collapse $J$, and the kink drift velocity $v_k$ along a dislocation line. Dislocation velocity does not depend on its length if the last is greater than the kink free path [2]

$$V = a\left(2v_k J\right)^{1/2}, \quad v_k = (D_k/kT)\,\tau ab, \quad D_k = \nu_D b^2 \exp(-W_m/kT) \qquad (1)$$

Here $a$ is the kink height, i.e. the distance between Peierls valleys; $\tau$ is the shear stress; $D_k$ is the kink diffusivity; $\nu_D$ is the Debye frequency; $W_m$ is the activation energy of the kink migration in the secondary Peierls relief.

The rate of a kink pair formation $J$ depends on the shear stress. The critical collapse-stable kink pair size is large under small stresses $t < 0.1\, t_p$; and the energy of a pair of kinks separated by a distance $x$ is described by the equation [3]

$$U(x) = 2U_k - tabx - a/x \qquad (2)$$

where $a$ is a constant of long-range elastic kink-antikink interaction; $U_k$ is the energy of a single kink. Plot $U(x)$ has the shape of a barrier (Fig. 1c) and a kink pair has to grow to the collapse-stable size before kinks could spread by a drift under applied stress $t$. Probability of the kink pair nucleation is [5]:

$$J \approx \frac{D_k U_k W_0''}{(kT)^{3/2}} \frac{(tab)^{3/4}}{ka^{1/4}} \exp\left(-\frac{2(U_k - \sqrt{taba})}{kT}\right) \qquad (3)$$

where $W_0'' = (d^2 W/dy^2)_{y=0}$; $\kappa$ is the dislocation line tension.

For higher stresses ($t_p > t > 0.1 t_p$) the short-range exponential interaction [6] predominates in the kink pair interaction

$$U(x) = 2U_k - tabx - (a^2/2)\sqrt{kW_0''}\exp(-\sqrt{W_0''/k}\,x) \qquad (4)$$

The critical collapse-stable kink pair size is of the order of a kink width, so it cannot be considered as a pair of isolated kinks, but as a whole excited state. It decays into the separate kink and antikink, giving rise to the dislocation motion. In this region $J$ depends exponentially on stress (as well as the dislocation velocity) [5]

$$J \approx J_0 \exp\left[-\left(2U_k - tab\sqrt{k/W''}\ln(pt_P/t)\right)/kT\right] \qquad (5)$$

The activation energy of the kink pair formation under low stresses ($t < 0.1 t_p$), $U_0 = 2(U_k - \sqrt{taba})$ depends weakly on stress as well as the dislocation velocity (1).

So, according to a theory, the potential relief of perfect crystal lattice determines two modes of dislocation mobility: the viscous motion (region III in Fig.2) and the thermoactivated motion (regions I and II in Fig.2). Thermoactivated regime of dislocation motion may proceed by two different modes of the formation of kink pairs of critical size: the diffusive, determining by Eq. (3) (region I in Fig. 2) and nondiffusive, determining by Eq. (5) (region II in Fig.2).

### Comparison of the theory predictions with experimental data on dislocation mobility

Let us compare the theory predictions with the experimental data obtained with 60° dislocations in Si [7, 8] (black circles in Fig. 2). Solid line shows schematically the theoretical dependence of dislocation velocity on stress. The velocities for thermoactivated modes were estimated for $U_k = 0.8$ eV and $W_m = 1.4$ eV. These values are average over data obtained in experiments [8–10]. The estimation of the Peierls stress [2] is $t_P \approx (p/2a)^3 (U_k^2/kb) \approx 900$ MPa. One can see the absence of the transition between the power (3) and exponential (5) modes of the thermoactivated motion. Particularly striking is the absence of the transitions to the viscous dislocation motion. Dislocations continue to move in the "low-stress" mode even under stresses $t > t_P$. It should be noted that using the uniaxial deformation of Si under a confining pressure, it was possible to obtain the stresses up to 3 GPa [11]. The contradiction cannot be resolved even under the assumption that the kink energy $U_k$ is equal to the total activation energy for the dislocation motion (2.2 eV). So, the dislocation motion in semiconductors under high stresses could not be adequately described in the context of current theory.

It is highly probable that new approaches should be invoked to resolve the paradox; for example, the model describing the formation of a kink pair as a result of evolution of nonlinear excitation of the breather type in the atomic subsystem of a crystal. The direct evidence of the existence of such excitations has been obtained for the magnetic subsystem of a crystal [12].

With low stresses the discrepancies are also observed for the dislocation motion in semiconductors [13, 14]. Experiments with Ge have revealed sharp decrease in the dislocation mobility with stresses approaching some low threshold [13]. The starting stresses for the dislocation motion have been revealed with the stress decrease in Si [14] (Fig. 2) and SiGe [15] crystals. Moreover, the change in activation energy for the dislocation motion with the rise of temperature both in Si and Ge has been found. The activation energy increases abruptly up to 4.0 eV in Si [16] and decreases down to 1.8 eV in Ge [17]. The asymmetry of the dislocation mobility has been found in Si and Ge [18]. The theory of dislocation motion in the perfect crystal made no prediction about these phenomena. The discrepancies can be partly resolved by taking into account the influence of point defects on the formation and motion of kinks.

## Modes of kink pair nucleation and spreading in the Peierls relief in presence of point defects

In crystals with point defects kink mobility may be determined by different modes of the thermoactivated motion. Two theoretical approaches have been developed. The first one [19] considers randomly distributed along the dislocation line barriers for the kink motion (Fig. 1c, curve 3). Both point defects near the dislocation and the dislocation core defects [20] may serve as the obstacles. In this approach the kink motion occurs in the random potential and is still described with Eq. (1) but with less kink diffusivity $D_k$. The rate of a kink pair formation $J$ decreases due to probability for the kink pair to collapse when it stays before the obstacle. This causes a strong decrease of dislocation velocity with a stress. The comparisons of experimental data on dislocation mobility in Si [21] and Ge [22, 23] with the theory [19] revealed that only qualitative agreement is possible.

Another approach is based on the cooperative effect of interaction of numerous point defects with a dislocation [24, 25]. Attachment of point defects to the dislocation core or their detachment due to kink motion causes the variation of the dislocation energy and therefore the specific step-like dependence of the energy of a kink pair on its size (Fig. 1c, curves 1, 4). The fluctuations of point defect density along the dislocation line causes the threshold dependence of the kink velocity on the stress at $\boldsymbol{t} > \boldsymbol{t}_0$

$$v_k = (D_k/kT)\cdot (\boldsymbol{t}-\boldsymbol{t}_0)\, ab \tag{6}$$

At $\boldsymbol{t} < \boldsymbol{t}_0$ the mean kink velocity becomes equal to zero (the kink quasi-localization occurs) [26]. Actually kinks do not stop completely, but the sublinear dependence of the path length $x$ on time $x \sim t^d$, $\boldsymbol{d} \leq 1$ takes place with the kink drift in the field of random forces.

On the other side, the inhomogeneous distribution of point defects across the dislocation causes the local change in dislocation energy (Fig. 1b) and the appearance of additional internal stresses favoring relaxation of the kink pair to its formation center (Fig. 1c, curve 1), that is the existence of starting stresses for the dislocation motion [27]. The nature of point defects interacting with dislocation and causing starting stresses in Si has been investigated in [28, 29].

## Experimental data on kink dynamics and estimates

First experimental estimations of the kink mobility have been obtained from the dependence of dislocation velocity on its length predicted by the theory of dislocation motion by Peierls mechanism [2]. If the dislocation length $L$ is greater than the mean kink free path $X$, the steady state dislocation velocity $V$ is determined by the production of the kink pair generation rate $J$ and kink drift velocity $v_k$ and does not depend on $L$ (Eq. (1)). However, if the dislocation segment length $L < X$, the dislocation velocity is proportional to its length $V = aJL$. The measurement of the critical dislocation length $L_c = X$, where change in the motion mode occurs, allows one to estimate mean kink drift velocity $v_k = V(X/a)$. It turned out that $X$ values for Ge and Si become rather small and could be measured only by *in situ* HVEM technique. Using Eq. (1) one may estimate kink diffusivity and the effective activation energy for the kink migration. The estimations have given a value of $W_m$=1.2–1.3 eV for silicon [9, 10] and $W_m$= 0.8–1.1 eV for germanium crystals [30].

To study experimentally the formation and motion of dislocation kinks in more details, the intermittent loading (IL) technique has been developed [8]. The method is based on the loading of a sample containing individual dislocations by a sequence of load pulses with the stress amplitude $t_i$. The duration of an individual pulse $t_i$ is comparable with a mean time of the dislocation displacement by one lattice parameter under conditions of the steady state motion $t_a=a/V_{st}$ where $V_{st}$ is a mean dislocation velocity under conventional static loading. The pulses are divided by 'pauses' with the duration $t_p$ when either the stress is not applied at all ($t_p$=0) or small enough stress $t_p$«$t_i$ of opposite sign with respect to $t_i$ is applied.

During the pulse stress action, in addition to thermodynamically equilibrium kinks, the extra kink pairs form and spread along the dislocation line. As the pulse separation goes on they become unstable and can collapse to the formation. By varying the duration of the pulses $t_i$ and the "pauses" $t_p$ and measuring the characteristics of dislocation distributions with respect to distances covered one may obtain information on the kinetics of the kink pair formation, spreading and relaxation [15, 31, 32].

Experiments with Si [8, 33] and SiGe [15, 34] revealed that dislocation displacements under IL diminish to zero with the pulse duration decrease (Fig. 3a, curves 2, 3). This result may be explained, in the frames of a model [27], considering the local decrease of dislocation energy in the Peierls valley due to interaction with mobile point defects (Figs. 1b and 1c, curve 1).

Unlike silicon, there are no starting stresses for the dislocation motion in low-doped Ge crystals and dislocation displacements do not drop to zero with small pulse durations (Fig. 3a, curve 1). However a sharp decrease in the dislocation mobility with lowering stress is observed [13]. We found, that dislocation velocities in Ge [31] and Si$_{1-x}$Ge$_x$ ($x$ = 0.055) [32, 34], may be described satisfactorily under low stresses by equation $V = ac_kv_k$ with $v_k$ being determined not with Eq. (1), but with Eq. (6). Figure 3b shows the dependences of mean dislocation glide distances in Si$_{1-x}$Ge$_x$ ($x$ = 0.055) and Si on the relative pulse separation. One can see that the decrease of dislocation glide distances is essentially nonlinear. Solid line in Fig. 3b, curve 1 shows the result of fitting of experimental data with the nonlinear kink drift in the field of random forces. One can see that experimental data are described well with the theory [24]. The dependence obtained with dislocations in Si (Fig. 3b, curve 2) may be explained qualitatively [32, 35] as a result of the immobilization of dislocations with the pulse separation increase due to interaction with mobile point defects.

The translational symmetry of a crystal lattice determines the secondary Peierls relief for the kink motion [2] (Fig. 1c, curve 2). The theory predicts a transition from the thermoactivated (1) at low stresses to the activationless viscous kink motion. Like in case of dislocations, nobody has revealed the viscous kink motion on dislocations in semiconductors yet.

It should be noted in conclusion that we have not considered the splitting of dislocations into partials. It was shown in [36] that dislocation splitting does not change the qualitative characteristics of the dislocation motion. Using HVEM it become possible to get direct image of the single kinks on the partial dislocations in silicon [37]. The migration energy of kinks on 90° partials was found to be $W_m$ = (1.24 ± 0.07) eV. The formation energy of a single kink is estimated to be $U_k$ = (0.73 ± 0.15) eV. These values are close to the estimations made for the whole 60° dislocations.

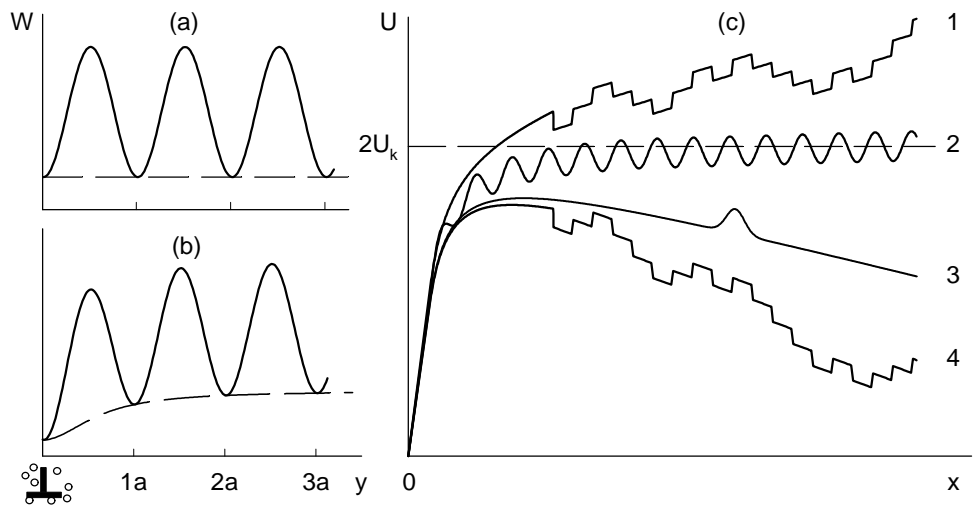

Figure 1. Peierls relief for dislocation motion in perfect crystal (a) and in crystal with point defects, decreasing the dislocation energy (b); free energy of a kink pair vs. its width in the perfect secondary Peierls relief, $t = 0$ (2); in the field of random forces, $t = 0$ (1) and $t > 0$ (4); in the random potential, $t > 0$ (3).

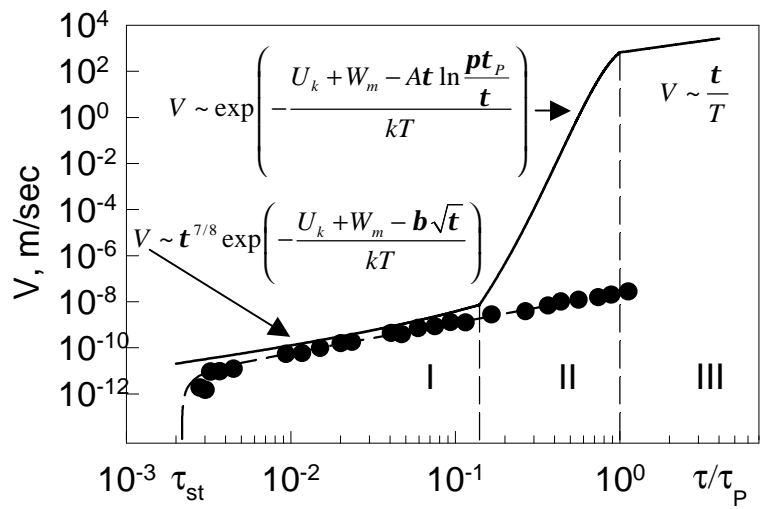

Figure 2. Stress dependence of dislocation velocity in Si: experimental data (black circles) and theoretical dependence (solid line).

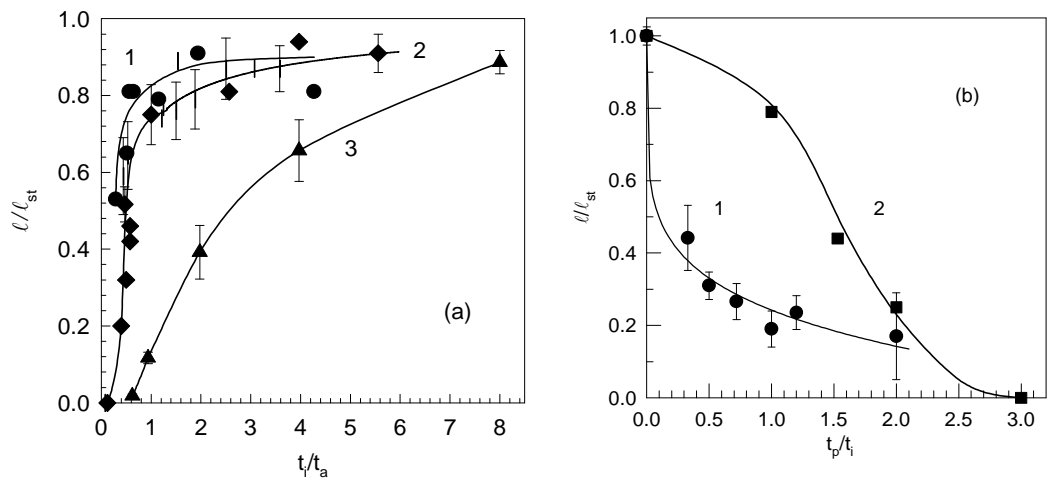

Figure 3. a – average 60° dislocation glide distances under IL normalized to the value with static loading in Ge (1), Si (2) and bulk $Si_{1-x}Ge_x$ ($x = 0.048$) (3) single crystals as a function of a relative pulse duration ($t_p = t_i$); b – average 60° dislocation displacement vs. relative pulse separation in bulk $Si_{1-x}Ge_x$ ($x = 0.055$) (1) and Si (2), $t_i = 30$ ms, $t_i/t_a = 0.83$, $\boldsymbol{t}_i$=15 MPa $T = 873$ K.